# Transfer Functions for the DAMA Experiments

P.A. Sturrock[a,*], J. Scargle[b] E. Fischbach[c], J.H. Jenkins[d], J. Nistor[c]

[a] Center for Space Science and Astrophysics, Stanford University, Stanford, CA 94305, USA
[b] NASA/Ames Research Center, MS 245-3, Moffett Field, CA 94035, USA
[c] Department of Physics, Purdue University, West Lafayette, IN 47907, USA
[d] Department of Nuclear Engineering, Texas A&M University, USA
College Station, TX 77843, USA
*Corresponding author. Tel +1 6507231438; fax +1 6507234840.
Email address: sturrock@stanford.edu

**Abstract**

Data acquired by the DAMA (DArk MAtter) Collaboration show strong evidence for an annual oscillation in the measured count rate with a phase compatible with that expected of dark matter in our galaxy. Analysis of their data (as reconstructed from figures in DAMA publications) by a likelihood procedure that takes account of data binning strengthens the evidence for a significant annual oscillation that the Collaboration obtained by means of Lomb-Scargle analysis. However, in view of the suggestion that DAMA signals may be due in part to beta-decays of $^{40}$K, it is important to know whether all features of the data are compatible only with the dark-matter hypothesis. We therefore examine what can and what cannot be revealed by the available DAMA dataset by computing the relevant amplitude transfer function (the ratio of the amplitude of the detected signal to that of an input signal) and the corresponding power transfer function. We find that the transfer functions drop off rapidly with frequency, virtually eliminating information concerning frequencies higher than ~3 year$^{-1}$. It has been reported that, for each bin, the count rate is normalized with respect to the associated 1-year average. We show that this procedure depresses the transfer functions for frequencies less than ~1 year$^{-1}$. Hence it is impossible to determine whether or not there are patterns in the basic DAMA measurements that are similar to oscillations found in certain beta-decay experiments in the frequency range 9 – 14 year$^{-1}$ (due apparently to internal solar rotation) and/or to oscillations with frequencies much less than 1 year$^{-1}$ (due apparently to solar r-mode oscillations). It will be necessary to examine the raw event data in order to extract all significant information from the DAMA experiments.

## 1 . Introduction

There is great interest in the possibility of detecting dark matter, the existence of which has been inferred from astrophysical data. As a result of the Earth's orbit around the Sun, the incidence of dark-matter particles on Earth is expected to have its maximum value at approximately June 2, when the Earth's orbital velocity is added to that of the Sun with respect to the Galaxy. To search for such an effect, the DAMA (DArk MAtter) Collaboration has operated a sequence of two experiments, referred to as DAMA/NaI and DAMA/LIBRA (Large sodium Iodide Bulk for RAre processes). These record signals from massive and highly radiopure sodium-iodide scintillators (100kg of Na(Tl) for DAMA/NaI and 250 kg for DAMA/LIBRA). The DAMA experiments, located at the Gran Sasso National Laboratory, have now been running for over 13 annual cycles with a cumulative exposure of over 1.17 ton-yr. Data published by the DAMA Collaboration present evidence (at a claimed confidence level of 8.9 σ) for an annual modulation with a maximum near June 2. [1-5]

We have recently undertaken an independent analysis of the DAMA data, which we have extracted from their various publications. [1-5] A power-spectrum analysis of this reconstructed dataset [6], using a likelihood power-spectrum procedure that is designed to take account of both experimental errors and bin durations [7,8], yields evidence for an annual oscillation stronger than evidence obtained by the DAMA Collaboration [3] using the Lomb-Scargle procedure [9,10]. The likelihood analysis also confirms that the phase of the maximum is near 0.4 (i.e. near June 1), which is compatible with that expected of dark matter in our galaxy.

However, other experiments aimed at detecting dark matter [11 - 15] do not support the claimed results of the DAMA Collaboration. Since potassium is an unavoidable contaminant of sodium, Pradler et al. [16,17] and Nistor et al. [18] have suggested that signals detected by DAMA may in fact be due to $^{40}$K beta-decays rather than to dark matter, but the DAMA Collaboration does not agree that this is a possibility [19].

In recent years, several investigators have found evidence that, for some (but not all) nuclides, the beta-decay rates are not constant. [20 - 26] Typically, the most prominent variation is an annual oscillation. However, we have also found evidence for oscillations in the frequency band 9 – 14 year$^{-1}$, which we attribute to the influence of the Sun's internal rotation [27 - 30], and to oscillations with frequencies less than 1 year$^{-1}$, which we attribute to r-mode oscillations in an inner tachocline separating the radiative zone from the core [31 - 33]. Hence a search for oscillations other than an annual oscillation may provide information relevant to the possible role of $^{40}$K.

In order to determine whether or not such oscillations might be detectable in the available DAMA data, we here examine the amplitude and power transfer functions, the ratio of the amplitude (or power) of a detected oscillation to the amplitude (or power) of an injected oscillation.

## 2 . The Transfer Functions for the DAMA Experiments

We define the *amplitude transfer function* as follows: For each relevant value of the frequency, we replace the data by an input modulation of unit amplitude and measure the output amplitude at that frequency, processing the data by whichever procedure is being used for the power-spectrum analysis. For each frequency, we repeat this procedure for a range of values of the phase of the input modulation. We define the amplitude transfer function as the mean value of the output modulation. We have applied this procedure to the DAMA data as listed in ref. [6], processing the data by a likelihood procedure that takes account of the start time and end time of each bin [7,8]. The result, smoothed by forming 7-point running means, is shown in Figure 1.

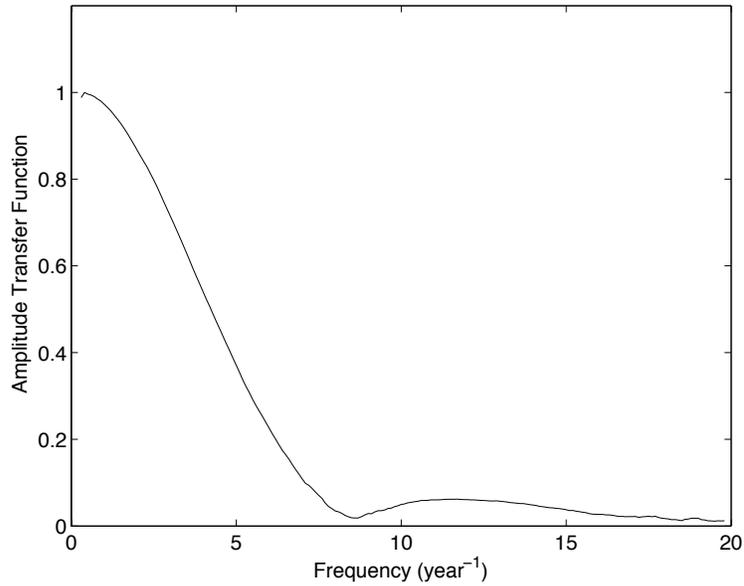

Figure 1. Smoothed and phase-averaged amplitude transfer function for the combined DAMA experiments, taking account of the binning of data.

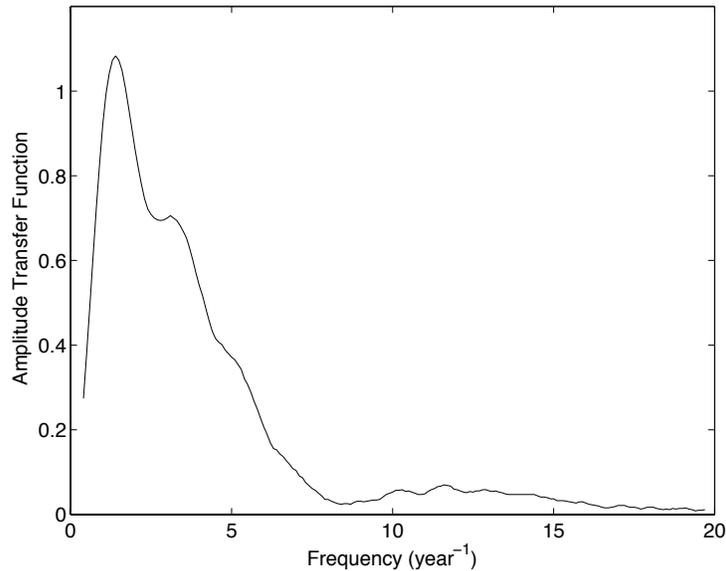

Figure 2. Smoothed and phase-averaged amplitude transfer function for the combined DAMA experiments, taking account of the binning of data and of mean subtraction.

However (according to Chang et al. [34]) for each bin, the DAMA Collaboration subtracts the annual mean count rate. It is well known that this operation forces the power to be zero at zero frequency, and thus leads to a drastic distortion of the low-frequency part of the power spectrum. Hence this procedure reduces the transfer function for low frequencies. The resulting form of the amplitude transfer function (again phase-averaged and smoothed by forming 7-point running means) is shown in Figure 2. (The curious fact that, for a small range of frequencies, the transfer function exceeds unity is found to be an unexpected consequence of normalizing measurements with respect to one-year averages.)

Since the significance of a peak in a power spectrum is usually derived from the power rather than the amplitude, it is interesting to study the power transfer function, which is simply the square of the amplitude transfer function. This function (phase-averaged and smoothed by forming 7-point running means) is shown in Figure 3.

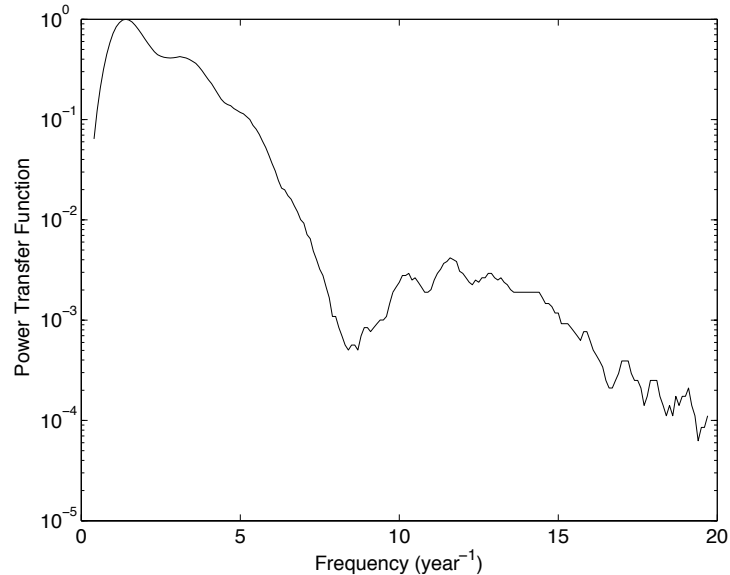

Figure 3. Smoothed and phase-averaged power transfer function for the combined DAMA experiments, taking account of the binning and of mean subtraction.

## 3 . Discussion

We see that the binning and mean-subtraction procedures that we understand to have been adopted by the DAMA Collaboration have severe adverse effects concerning time-series analysis. We see from Figure 3 that, as a result of the binning, the powers of oscillations with frequencies greater than about 5 year$^{-1}$ are reduced by a factor of 100 or more.

The frequency range 9 – 14 year$^{-1}$ is of particular interest, since this is the range of synodic solar rotation frequencies that one finds in analyses of certain records of anomalous beta decays. [27 - 30] We see from Figure 3 that, for this frequency range, the power is reduced by a factor of about 500. Hence any such oscillations in the signals registered by the DAMA experiments would be effectively eliminated by the DAMA binning procedure.

Some experiments show evidence of oscillations with frequencies less than 1 year$^{-1}$, which may be attributable to r-mode oscillations in a tachocline separating the solar core from the radiative zone. [31 - 33] For instance, analysis of data concerning the decay of $^{90}$Sr shows evidence of such oscillations at 0.77 year$^{-1}$ (with power 19), at 0.47 year$^{-1}$ (power 57), and at 0.26 year$^{-1}$ (power 73). [33] We find from the data leading to Figures 2 and 3 that the power at 0.26 year$^{-1}$ would be reduced by a factor of 100. Even if the amplitude of the oscillation at this frequency were as great as that of the annual oscillation, which has a power of 37.10 [11], it would appear to be insignificant (with power 0.4) if analyzed according to the procedure adopted by the DAMA Collaboration.

A recent analysis of radon data yields evidence of diurnal oscillations. [30] This oscillation may be associated with the specific geometrical configuration of that experiment. Nevertheless, one cannot at this stage of knowledge rule out the possibility that the DAMA experiment may exhibit a similar effect.

In order to fully understand the significance of the annual oscillation detected by the DAMA experiments, it will be essential to carry out a time-series analysis that has a transfer function that is as close to unity as possible over as wide a frequency range as possible. This could best be achieved by releasing the raw data that presumably consists of the precise timing of each event registered by the NaI detectors. If a search over a wide frequency band shows no evidence for any oscillation other than the annual oscillation, that would strengthen the case that DAMA is detecting a cosmological process. On the other hand, if such a search yields evidence of oscillations that have a solar interpretation, one would then need to determine whether DAMA is responding only to solar influences, or to both solar and cosmic influences.

**Postscript**

We have just become aware of a recent summary article by the DAMA Collaboration (arXiv:1306.1411). This article contains power-spectrum calculations that are based on the Lomb-Scargle procedure with modifications that are intended to take account of binning and error estimates. The DAMA Collaboration does not justify these modifications nor give a reference to a book or article where the proposed modifications are derived. Furthermore, we find that results which the Collaboration obtains using these modifications do not agree with the results of our calculations that use a likelihood procedure designed to take account of both binning and error estimates. For instance, Figure 3 of their article shows the result of applying their formalism to the combined DAMA/NaI and DAMA/LIBRA datasets. For the annual modulation, it yields a power of 21.7, which appears to be indistinguishable from the value given in ref. [3]. We have found that a Lomb-Scargle analysis of the reconstructed DAMA data (DAMA/NaI and DAMA/LIBRA) yields a peak power of 19.2. Hence the modifications appear to have only a minor effect on the power estimates.

On the other hand, we have found that a likelihood procedure that is designed to take account of binning and of the experimental errors leads to a peak power of 37.1. [11] Hence the procedure we advocate for taking account of binning and error estimates is much more favorable for the DAMA project than the procedure that they have used.

The new DAMA article shows the result of their power spectrum analysis for frequencies up to 22 year$^{-1}$ (0.06 day$^{-1}$). However the power transfer function is a limiting factor whichever procedure one adopts for power-spectrum calculations. It is therefore not surprising that the DAMA power spectrum shows no indication of any periodicity other than the annual periodicity. It remains an open question as to whether power-spectrum analysis of unbinned data will or will not reveal any additional modulations.